\documentclass[referee]{aastex}

\begin{document}

\title{The origin and propagation of variability in the outflows of \\ long
duration gamma-ray bursts}

\shorttitle{Variability in GRBs}

\author{Brian J. Morsony\altaffilmark{1}, Davide Lazzati\altaffilmark{2}
and Mitchell C. Begelman\altaffilmark{3,4}} \shortauthors{Morsony et
al.}

\altaffiltext{1}{Department of Astronomy, University of
Wisconsin-Madison, 3321 Sterling Hall, 475 N. Charter Street, Madison WI
53706-1582}

\altaffiltext{2}{Department of Physics, NC State University, 2401
Stinson Drive, Raleigh, NC 27695-8202}

\altaffiltext{3}{JILA, University of Colorado, 440 UCB, Boulder, CO
80309-0440}

\altaffiltext{4}{University of Colorado, Department of Astrophysical and
Planetary Sciences, 389 UCB, Boulder, CO 80309-0389}

\begin{abstract} 
  We present the results of hydrodynamical simulations of gamma-ray
  burst jets propagating through their stellar progenitor material and
  subsequently through the surrounding circumstellar medium. We
  consider both jets that are injected with constant properties in the
  center of the star and jets injected with a variable luminosity. We
  show that the variability properties of the jet outside the star are
  a combination of the variability injected 
  at the base of the jet 
  and the variability caused by the jet propagation through the
  star. Comparing power spectra for the two cases shows that the
  variability injected by the engine is preserved even if the jet is
  heavily shocked inside the star. Such shocking produces additional
  variability at long time scales, of order several seconds. Within
  the limited number of progenitors and jets investigated, our
  findings suggest that the broad pulses of several seconds duration
  typically observed in gamma-ray bursts are due to the interaction of
  the jet with the progenitor, while the short-timescale variability,
  characterized by fluctuations on time scales of milliseconds, has to
  be injected at the base of the jet. Studying the properties of the
  fast variability in GRBs may therefore provide clues to the nature
  of the inner engine and the mechanisms of energy extraction from it.
\end{abstract}


\section{Introduction}

Gamma-ray bursts (GRBs) are extremely bright and highly variable
sources of gamma-ray radiation. Their isotropic equivalent emitted
energy can be as large as $10^{55}$~erg (GRB080916C; Abdo et
al. 2009). GRBs last between a fraction of a second and several
thousand seconds and can be divided in two classes based on their
duration and spectral characteristics (Kouveliotou et al. 1993). Short
bursts last less than 2~s, while long bursts last between 2 and
several thousand seconds. Even though some bursts seem hard to
classify within this simple scheme (e.g., GRB~060505 and GRB~060614;
Fynbo et al. 2006, Gal-Yam et al. 2006, Della Valle et al. 2006;
GRB~060121, de Ugarte Postigo et al. 2006) it is widely believed that
a substantial fraction of, if not all, long-duration GRBs are
associated with the death of massive, rapidly spinning stars (Woosley
1993; MacFadyen \& Woosley 1999; Stanek et al. 2003; Hjorth et
al. 2003; Woosley \& Bloom 2006).

Within the overall duration of the prompt phase, fast variability is
commonly observed, with the shortest spikes lasting only a fraction of
a millisecond (Schaefer \& Walker 1999, Walker et al. 2000).
Characterizing the burst variability has proved challenging, with each
GRB seeming to have its own individual pattern. This is particularly
frustrating since the variability can in principle carry information
on the workings of the central engine, the energy dissipation
processes, and the radiation mechanisms involved in the release of the
burst emission. Fenimore \& Ramirez-Ruiz (1999) showed that the
variability time scale of GRB light curves does not evolve from the
beginning to the end of the prompt emission. Their work placed a
strong constraint on the dissipation process and proved that the
variability in the GRB prompt emission is not due to the interaction
of the fireball with the external medium (see Dermer et al. 1999; 2000
for an alternative interpretation).  Beloborodov et al. (1998, 2000)
analyzed the composite power spectrum of BATSE light curves, finding
that it is well described by a power-law of the shape $PDS(f)\propto
f^{-5/3}$, reminiscent of the Kolmogorov spectrum of fully developed
turbulence.

The discovery of the association of GRBs with massive stars puts burst
variability in a new light. There are two possible sources of
variability in the outflow: the engine itself (MacFadyen \& Woosley
1999; Aloy et al. 2000; Ouyed et al. 2003; Proga et al. 2003; McKinney
\& Narayan 2007; McKinney \& Blandford 2009) and the interaction of
the flow with the progenitor star material (Aloy et al. 2002; Gomez \&
Hardee 2004; Aloy \& Obergaulinger 2007; Morsony et al. 2007). How
these two sources of variability combine and interact to give rise to
the variability in the light curve is unknown but of fundamental
importance if any conclusion is to be drawn from the temporal
properties of GRB light curves.

In this paper we present the results of 2D axisymmetric simulations of the
propagation of baryonic GRB jets through their progenitor star
material.  We show simulations in which the engine is either constant
or variable and compare the structure of the jets as they emerge from
the progenitor star. This paper is organized as follows: in \S~2 we
describe our simulations, in \S~3 we describe our results, and in \S~4
we discuss their potential implications.

\section{The simulations and the injected variability}

\begin{figure} 
\plotone{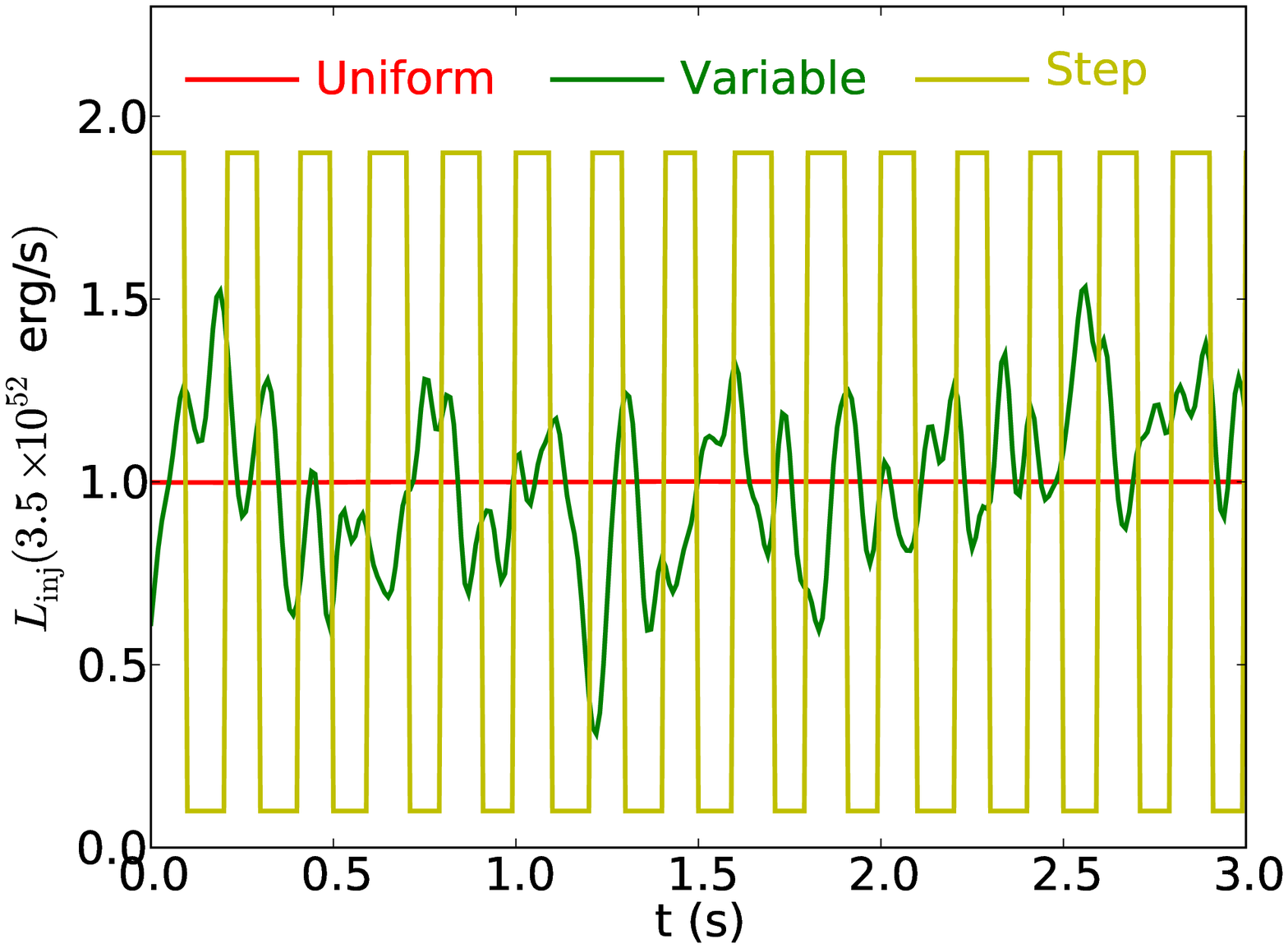} 
\caption{{Luminosity injected by the central engine for the four
    simulations described in the text. The figure shows only a 3
    second zoom of the simulations that last a total time of 50
    seconds. The luminosity is normalized to the average isotropic
    equivalent luminosity of $L_{\{\rm{inj,
        iso}\}}=3.5\times10^{52}$~erg/s.  The injected luminosity for
    the variable entropy and variable baryon load simulations is the
    same.}
\label{fig:lum}} 
\end{figure}

We present the results of four simulations, each of which has the same
progenitor star and average jet properties, but with different
injected luminosity histories. The first simulation has uniform
injected luminosity and is analogous to that already presented in
Morsony et al.  (2007, hereafter MLB07) and Lazzati et al. (2009
hereafter LMB09). We call this simulation {\it uniform}. The second
and third simulations ({\it variable entropy} and {\it variable baryon
 load}) represent an inner engine that releases a luminosity that varies
in time. The time history was simulated as a random series of numbers
convolved with a Gaussian, to yield a flat power spectrum (white
noise) out to a cutoff at $\sim0.1$s. 
This represents a randomly varying energy input, with a cutoff such that all the variability is well resolved.
The energy input in a real GRB could, of course, have a different power spectrum, but for our purpose of determining how the variability is modified during propagation, this input is an adequate example.
The difference between the
simulation that we call {\it variable entropy} and the one that we
call {\it variable baryon load} is that in the former the luminosity
is varied by changing the entropy $\eta=L/\dot{m}c^2$ holding
$\dot{m}$ constant, while in the latter the luminosity is changed by
changing the baryon load ${\dot{m}}$. Finally, a simulation was run
with an on-off engine with a period of 0.2s (0.1s on, 0.1s off), in
which the luminosity was changed by changing the baryon load ${\dot{m}}$. We identify
this last simulation as {\it step}. 
The represents an extreme case of strong, fast variability but on a time scale that is well resolved in our simulation.
A 3s section of the injected luminosity for the four simulations 
is shown in Figure~\ref{fig:lum}. 

\begin{figure}
\plotone{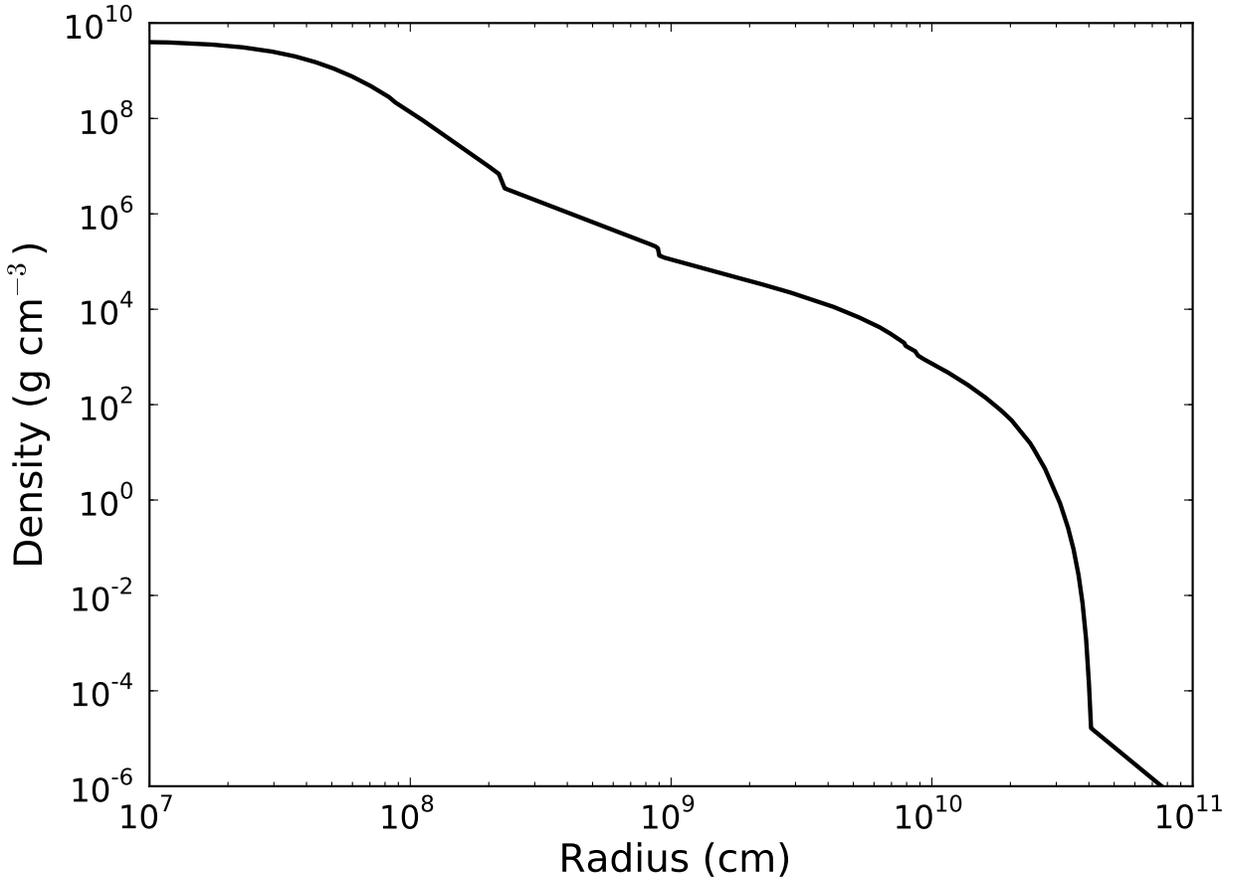}
\caption{{Density profile of the model progenitor star 16TI adopted in
    our simulations. From Woosley \& Heger (2006).}
\label{fig:16ti}}
\end{figure}

In all simulations, the progenitor star is model
16TI from Woosley \& Heger (2006, see its density profile in
Fig.~\ref{fig:16ti}), the jet is injected at a distance
$R_0=10^{9}$~cm from the center of the star and has an initial opening
angle $\theta_0=10^\circ$ and an initial Lorentz factor
$\Gamma_0=5$. In all simulations self-gravity is neglected, and
therefore the star expands slightly under the effect of its internal
pressure during the 50 seconds of our simulation. The expansion is
however negligible (less then 1\% of the radius) and does not affect
the dynamics of the jet. In all simulations the surrounding medium is
uniform, with a density $\rho=10^{-9}$~g~cm${-3}$ (for the big box
simulation discussed below, the density was reduced to
$\rho=10^{-13}$~g~cm${-3}$, LBM09). Further simulations
will be performed to study the effect, if any, of a wind environment
on the jet evolution outside the stellar progenitor.

\begin{figure} 
\plotone{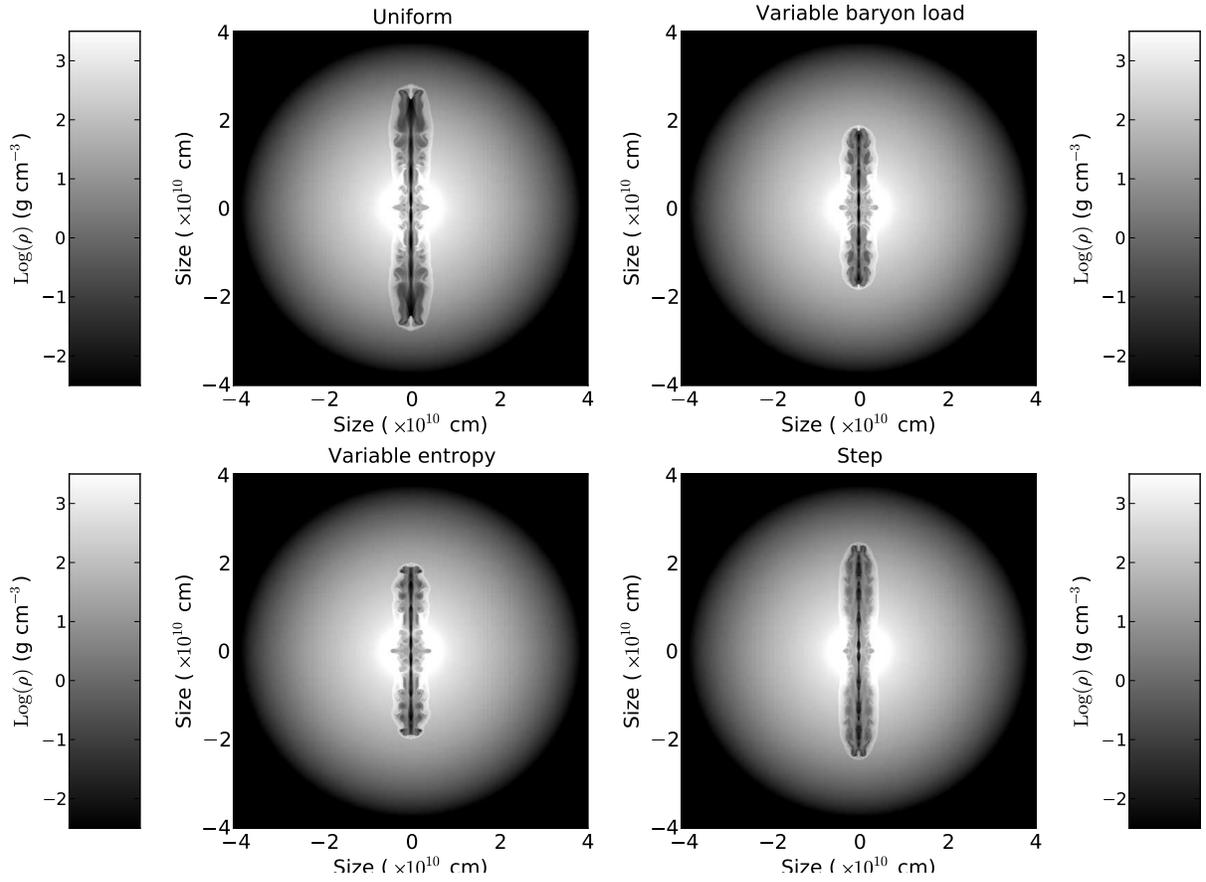} 
\caption{{Density images of our four simulations 5.5 seconds after the
    onset of the central engine.}
\label{fig:panels}}
\end{figure}

The simulations were performed in 2D cylindrical coordinates with the adaptive mesh refinement (AMR)
special relativistic hydrodynamic code FLASH (Fryxell et al. 2000; see
MLB07 for extensive testing of the special relativistic module), as
modified by the authors (MLB07). The simulations were carried out for
50 seconds for a maximum number of refinements of 13 in the stellar
core, corresponding to a maximum resolution of
$7.8125\times10^{6}$~cm.  Outside $5\times10^{9}$~cm from the center
of the star, the maximum resolution is $3.125\times10^{7}$~cm,
corresponding to 11 levels of refinement.  This is 4 times the
resolution used in MLB07 and LMB09. 
The temporal resolution of these simulations is $1/100$th of a second.
Figure~\ref{fig:panels} shows
logarithmic density contours for the four simulations at $t=5.5$~s. In
all panels, comoving density is shown.  In all cases, a very narrow
low-density jet is visible surrounded by a highly structured,
turbulent cocoon of shocked stellar material.

\section{Results}

\subsection{Effect of a variable input on the jet propagation inside the
  star}

We first concentrate on the way in which the jets propagate through
the material of their stellar progenitor. Extensive work on this topic
has been done in the past (MacFadyen \& Woosley 1999; Aloy et
al. 2000; MacFadyen et al.  2001; Zhang et al. 2003, 2004; MLB07,
Mizuta \& Aloy 2009) but only in one case (Aloy et al. 2000) were
variable engines considered. Aloy et al. concluded that a variable
engine has a faster propagation time through the stellar progenitor.

\begin{figure} 
\plotone{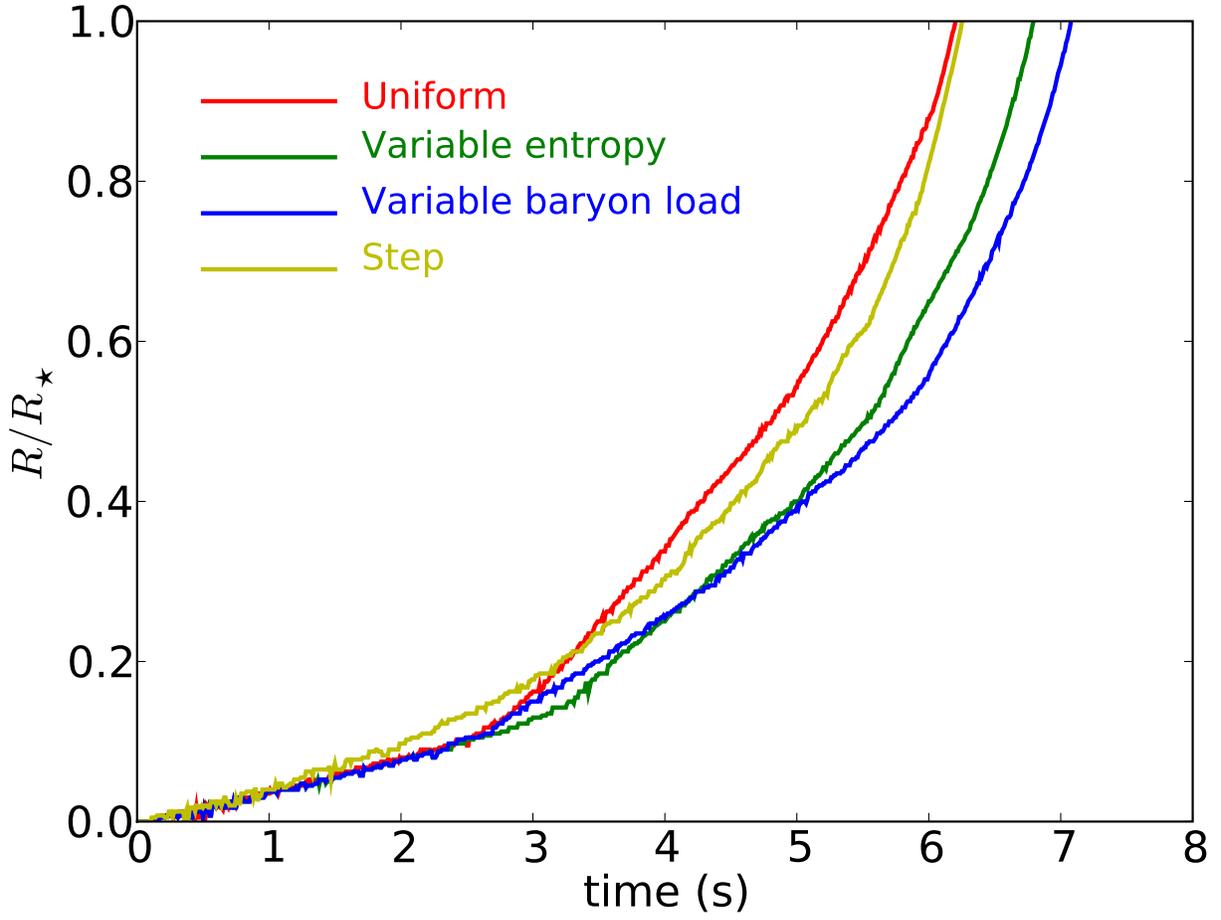} 
\caption{{Propagation of the jet head through the progenitor star for
    the four simulations described in the text.}
\label{fig:jh}} 
\end{figure}

Figure~\ref{fig:jh} shows the propagation of the jet head through the
progenitor star for our four simulations. The jet head was identified
as the location of the bow shock where the outward velocity of the
material exceeds 1\% of the speed of light along the jet
axis. Contrary to previous results, we find that the shortest
propagation time is observed for a uniform injected luminosity, with
the head of the jet breaking out of the star's surface at about 6.2s,
corresponding to an average velocity $v_{\rm{jh}}=0.21c$. The {\it
  step} simulation has a slightly longer propagation time of 6.25~s,
while the {\it variable entropy} and {\it variable baryon load}
simulations have longer breakout times of 6.8 and 7.1~s,
respectively. It is not straightforward to understand the origin of
the difference between our findings and those of Aloy et al. (2000).
One possibility is that thanks to the AMR capabilities of the FLASH
code, the resolution of our simulation is higher, especially as the
jet approaches the stellar surface. On the other hand,
Figure~\ref{fig:jh} shows that the effect of the engine variability is
not simple. In the core of the star, out to approximately 20 per cent
of its radius, the {\it uniform}, {\it variable entropy}, and {\it 
variable baryon load} jets propagate at the same speed, while the
{\it step} simulation (the one most similar to the Aloy et al. 2000
set up) is ahead. At larger radii, the behaviors differentiate, with
the uniform simulation showing the fastest velocity and reaching the
stellar surface first. It is therefore possible that the difference
between our result and the result of Aloy et al. (2000) is due to a
different structure of the progenitor star. Alternatively, the
difference may be due to the different prescription used in Aloy et
al. (2000) for the jet injection: while we injected an outflow with
net momentum as a boundary condition, Aloy et al. (2000) injected
energy with no net momentum in a conical region in the core of the
star
, allowing a jet to develop rather than be injected.
The lower boundary of the grid used by Aloy et al. (2000) is also much farther inside the star, at $2\times10^{7}$~cm rather than $10^{9}$~cm.  Differences in jet propagation may arise in this inner region that our simulations are unable to capture.
Finally, it is possible that the speed of the jet head depends on
the duration of the on and off phases or the frequency at which the
variability is injected. The difference between our result and Aloy et
al. (2000) may simply be due to the fact that we investigate
variability at different frequencies.

\begin{figure} 
\plotone{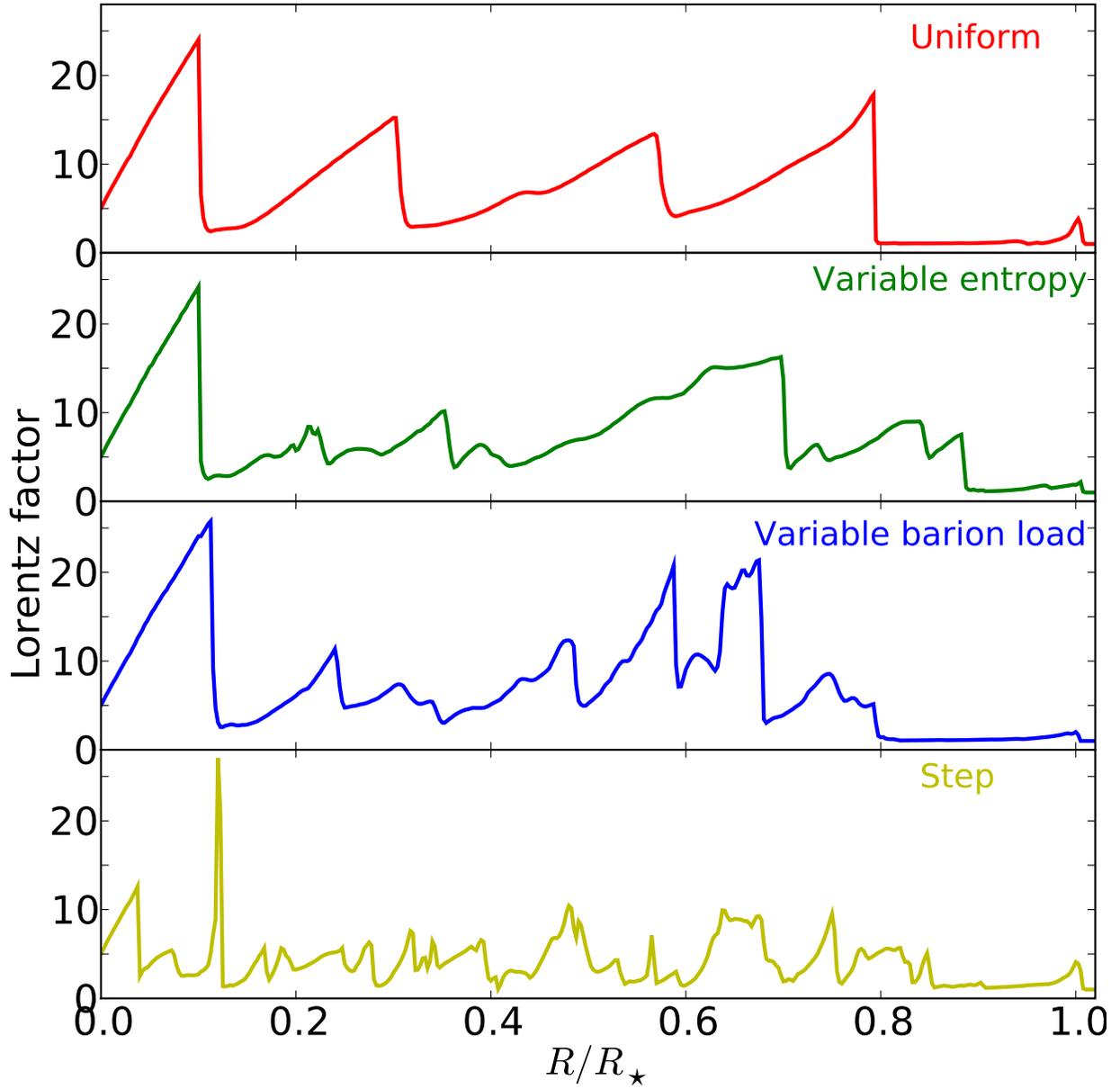}
\caption{{The dynamical conditions of the jet inside the star at the
    breakout time. The four panels show the Lorentz factor along the
    jet axis for the four simulations described in the text.}
\label{fig:sh}} 
\end{figure}

The complexity and the sensitivity to details of the interaction of
the jet head with the star is also shown in Figure~\ref{fig:sh}. The
figure shows the Lorentz factor of the jet material along the jet axis
at the moment of breakout for the four simulations. The {\it uniform}
simulation has the most regular structure, with strong recollimation
shocks separated by acceleration phases in which the Lorentz factor
grows linearly with radius. All the {\it variable} simulations show a
more complex evolution of the Lorentz factor, with milder but more
numerous shocks. In particular the {\it step} simulation shows a very
complex evolution of the Lorentz factor in which strong recollimation
shocks are absent. This is probably due to the fact that a strong
shock cannot be long lived since it is lost when it propagates from a
low luminosity phase to a high luminosity one.

\subsection{Variability properties of the jets outside the star \label{section3.2}}

\begin{figure} 
\plotone{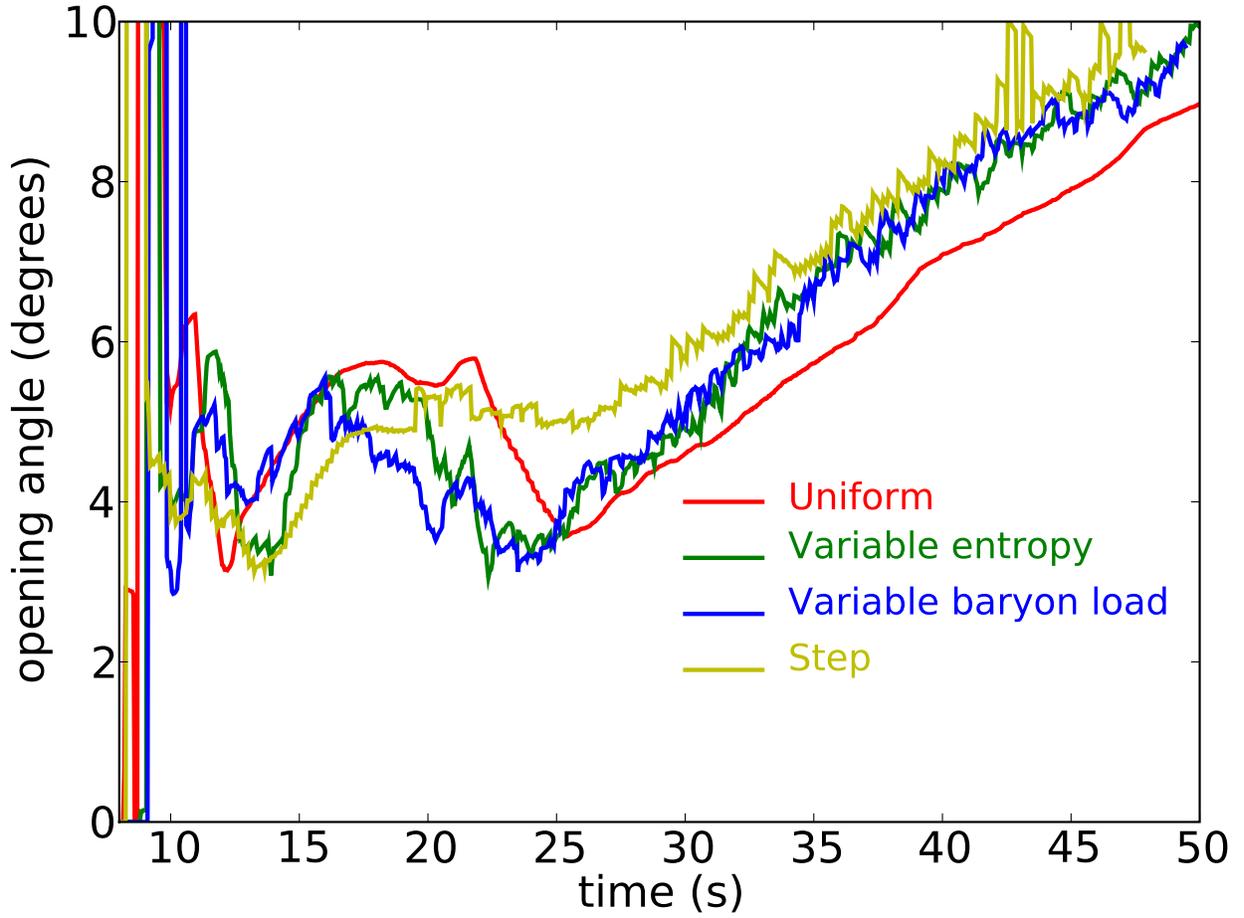} 
\caption{{Jet opening angle as a function of time for the four
    simulations described in the text. The jet opening angle is
    measured at a distance of $10^{11}$~cm from the center of the
    progenitor star. Despite differences in the details, the
    variability injected by the central engine does not affect the
    overall evolution of the jet opening angle.}
\label{fig:oa}} 
\end{figure}

While the propagation of the jet inside the progenitor star is
strongly dependent on details, the jet emerging at the surface of the
star and propagating out of it is fairly insensitive to those same
details.  Figure~\ref{fig:oa} shows the evolution of the opening angle
of the jet measured at a distance of $R=10^{11}$~cm from the center of
the progenitor star (or at 2.5 stellar radii). The opening angle is
measured as the geometric angle of the plasma parcel moving at
$\Gamma\ge10$ that is located farthest from the jet axis. As described
in MLB07, the opening angle is initially very large, due to the
quasi-isotropic cocoon material (see also Lazzati et al. 2010). Such a
phase, visible in Figure~\ref{fig:oa} as the spike in angle at
$t\sim10$~s, is followed by the shocked jet phase, in which the
opening angle of the jet is roughly constant and about half the size
at injection. This is the phase in which the jet interaction with the
progenitor star is the strongest and lasts, in all cases,
approximately 10 to 15 seconds. After that, the interaction with the
star becomes weak, and the jet gradually expands to its injection
opening angle\footnote{Note that if an injection opening angle is not 
assumed, such as in Aloy et al. 2000, the asymptotic opening angle 
is set by the jet and progenitor properties.}
$\theta_0=10^\circ$. Since the jet is injected with a moderate lorentz
factor $\Gamma_0=5$, the opening angle eventually expands to
$\theta_{\rm{lim}}=\theta_0+1/\Gamma_0\sim21\degr$ (MLB07).

Even though minor differences are present, all four simulations follow
the same qualitative trend, with the three phases clearly identified
and lasting approximately the same time, and with approximately the same opening angle evolution. 
This indicates that the structure of the jet that emerges from the progenitor star is similar regardless of the short-timescale variability of the energy injection.

\subsection{Light-power curves}

Another way in which the structure of the jet outside the star can be
investigated is through the calculation of light-power curves. The
light-power curve at a radius $R$ is computed as:
\begin{equation}
L_R(t)=c \int_{\Sigma_R} d\sigma \left[(4p+\rho c^2) \Gamma^2 -\rho
  c^2 \Gamma\right]\delta^2
\label{eq:lum} 
\end{equation} 
where $\Sigma_R$ is a spherical surface of radius $R$ centered on the
GRB engine, $p$ and $\rho$ are the pressure and comoving density of
the jet, respectively, and $\delta=[\Gamma(1-\beta\cos\theta)]^{-1}$
is the Doppler factor. The luminosity in Eq.~\ref{eq:lum} is formally
the luminosity that the observer would detect if all the energy were
released as radiation at the radius $R$ with a 100\% efficiency. The
light-power curves have properties of light curves, since they contain
the Doppler factor to beam photons in the direction of motion, and of
power-curves, since they assume 100\% efficiency and are computed at a
fixed radius.  In addition, we included in the computation only
material that has a minimum Lorentz factor
$\Gamma_{\infty,\rm{min}}=10$, since slower material is not supposed
to contribute to the keV-MeV prompt light curve of GRBs.  

Using the light power curve as a proxy for the light curve is a rough
approximation of the processes that lead to the generation of the
light curve in a GRB outflow.  However, given the impossibility of
following the propagation of the jet in 2D out to the radiative
radius, an assumption has to be made. The rationale behind our choice
is that the brightest spikes in the light curve are produced by the
most energetic parts of the outflow, and the duration of a spike
should be correlated to the radial thickness of the energetic part of
the outflow. Such close connection is true for the photospheric
scenario for the prompt emission (Rees \& Meszaros 2005; LBM09) and
for the internal shocks scenario, at least in first approximation
(Kobayashi et al. 2007). In other words, we believe our light power
curves have the right number of pulses with correct durations. If,
however, we analyze the details of the light curve of individual
pulses, our approximation is likely to break down, since the spectral
evolution is completely missed in our approximation (see, e.g., Daigne
\& Mochkovitch 2002). In addition, it has been shown that second
generation shells - shells produced by collisions between other shells
- are complex and would produce more complex pulses than original
shells (Mimica et al. 2005, 2007). So, we expect that our
approximation does miss some of the fine structure in the light curves,
but we also expect the overall qualitative behavior to be accurate.

\begin{figure} 
\plotone{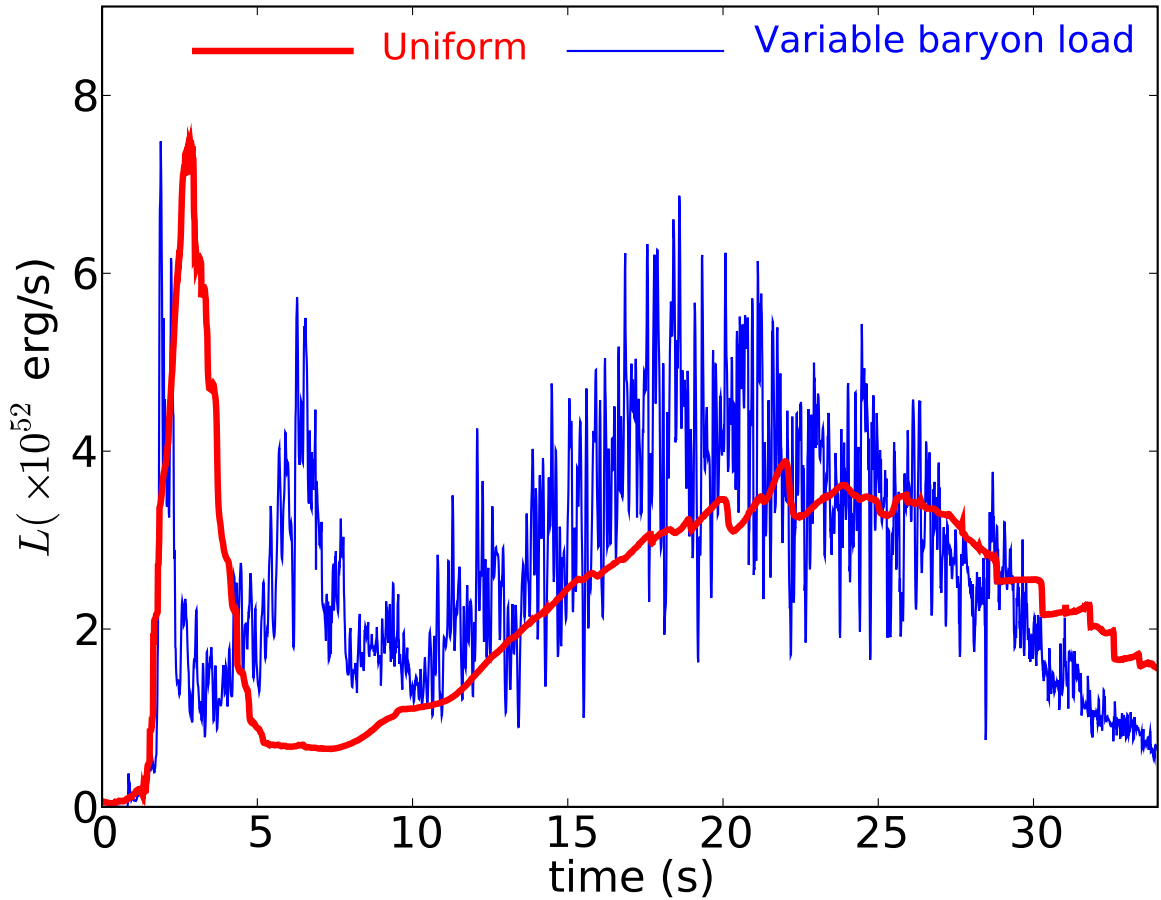} 
\caption{{Comparison between the light-power curve of the {\it uniform}
    simulation (thick red line) and the light-power curve of the {\it
      variable baryon load} simulation (thin blue line). In both
    cases, time is measured from the instant at which relativistic
    matter is first detected. The two curves show analogous long-term
    behavior, with an initial spike followed by a broad peak lasting
    approximately 20 seconds.  The light-power curve for the variable
    simulation also displays high-frequency variability at all times.}
 \label{fig:lc1}}
\end{figure}

Figure~\ref{fig:lc1} shows the light-power curve computed at a radius
$R=2.5\times10^{11}$~cm (the edge of our simulation box\footnote{To
  make sure the light-power curves are not affected by edge effects,
  we computed also light power curves at a radius
  $R=2\times10^{11}$~cm, obtaining fully consistent results}.) for the
{\it uniform} and the {\it variable baryon load} simulations. Several
important conclusions can be drawn from the figure. First, even if the
central engine releases a constant luminosity, the power curve shows
marked variability as a result of the interaction between the jet and
the progenitor star. This is not a new result (MacFadyen \& Woosley
1999; Aloy et al. 2000, 2002; Zhang et al. 2003, 2004; MLB07;
LMB09). The comparison between the two curves shows, however, a more
interesting and new conclusion. The characteristics of the long
time-scale variability are similar in the two light curves, with an
initial phase characterized by peaks with a width of several seconds
followed by a broad bump with a duration of approximately 20
seconds. Some differences are however apparent, with a second peak
appearing in the {\it variable baryon load} light power curve that is
not observed in the {\it uniform} simulation.

\begin{figure} 
\plotone{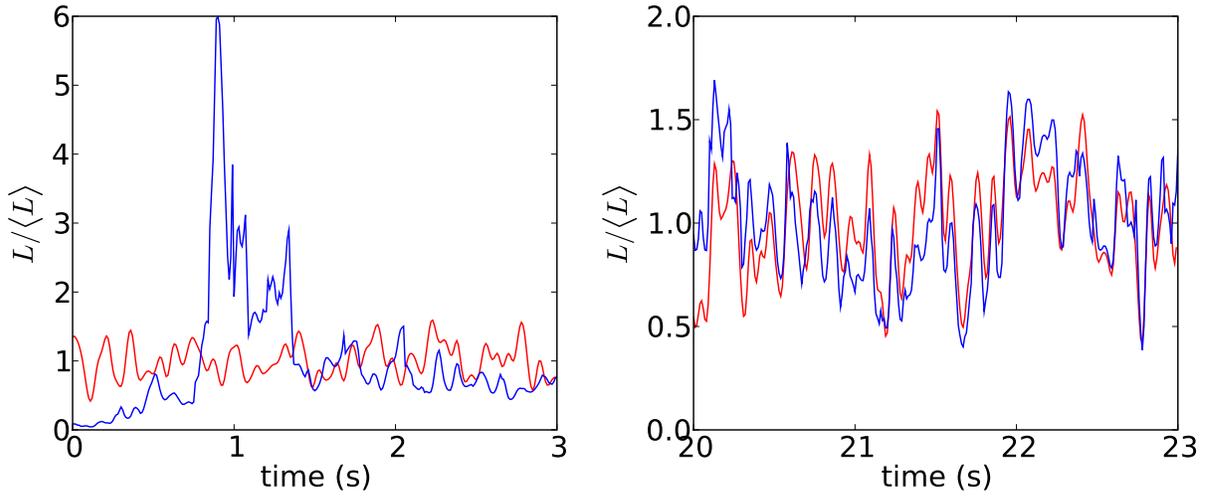} 
\caption{{Comparison between the variability injected by the inner
    engine (red line) and the one observed in the power curve at a
    radius of $R=2.5\times10^{11}$~cm. The left panel shows the
    initial 3 seconds of the light-power curve, while the right panel shows
    a central section of 3 seconds, during a phase in which the light
    curve is still deeply affected by the jet interaction with the
    progenitor star (cfr. Figure~\ref{fig:lc1}). While the variability
    of the two curves in the left panel is almost uncorrelated, a
    strong correlation is observed in the right panel.}
\label{fig:lc2}}
\end{figure}

Another interesting comparison is presented in Figure~\ref{fig:lc2},
which shows the luminosity injected at the base of the jet (red line)
compared to the light-power curve measured well outside the star at
radius $R=2.5\times10^{11}$~cm for the same {\it variable baryon load}
simulations shown in Fig.~\ref{fig:lc1}. The injected curve has been
shifted in time by adding the light propagation time from the inner
boundary of the simulation out to the radius at which the light power
curves have been computed. In this way, the jet propagation time was
removed. The left panel shows the comparison during the first three
seconds of the light-power curve while the right panel shows the same
comparison at a later time, still during the shocked jet phase
(according to the definition of MLB07). In the left panel, fast
variability is observed in the light-power curve, but it is almost
completely uncorrelated with the injected variability at the base of
the jet. On the other hand, at the later time shown in the right
panel, the variability of the engine and that observed in the
light-power curve are almost identical. A further analysis shows that
the strong correlation appears at about $\sim4$~s from the trigger and
lasts until $t=30$~s, the beginning of the unshocked jet phase
(MLB07). The same analysis was performed with the {\it variable 
entropy} and the {\it step simulations}. We found that in both cases
the strong correlation is observed. However, the {\it variable 
entropy} light power curve is sometimes shifted in time, likely due
to different propagation times of shells with different entropy and
therefore different asymptotic Lorentz factor. In the {\it step}
simulations, instead, the strong correlation is observed at all times,
likely due to the unphysically strong character of the fluctuations
injected in that model.

This suggests that there are two sub-phases within the shocked-jet
phase. An initial short phase, lasting a few seconds, has the
capability of modifying the variability injected by the inner engine,
while a second phase, lasting several tens of seconds until the end of
the shocked-jet phase, leaves the engine variability unaffected. A
closer inspection of our simulations reveals that the initial short
phase is characterized by perpendicular shocks, propagating along the
jet similarly to internal shocks, while the latter phase is
characterized by tangential shocks that propagate from the sides to
the jet axis and vice versa. The fact that the variability injected by
the central engine is preserved during the shocked jet phase comes as
a surprise. Both Zhang et al. (2003) and MLB07 had predicted, based on
simulations of constant engines, that the variability injected by the
central engine, if any, would have been visible only ``at very late
times when the star has exploded'' (Zhang et al. 2003). We find here,
from the analysis of a simulation with a variable inner engine, that
the transverse collimation shocks are not able to erase the
short-timescale (fraction of a second) variability injected at the
base of the jet by the inner engine. This finding makes us much more
optimistic that the characteristics of the central engine can be
studied by observing the light curves of GRBs. Further analysis is
however needed to confirm the robustness of this result for different
progenitors, injected luminosities, and jet properties.

The inner engine variability is also visible during the unshocked jet
phase, as described in MLB07. This final phase, however, is much
dimmer than the shocked jet phase and it is unclear whether it could
be observed at all in GRB light curves.

\subsection{Polar jet structure}

The computation of light-power curves allows us to compare the amount
of energy and the peak luminosity as a function of the off-axis angle
$\theta_o$. Previous studies on the distribution of energy in the
fireball with respect to $\theta_o$ have resulted in controversial
results. Zhang et al. (2004) found a somewhat universal slope
$dE/d\Omega\propto\theta^{-3}$, while MLB07 and Mizuta \& Aloy (2009)
found that the slope depends on the properties of the stellar
progenitor, and on the limiting Lorentz factor $\Gamma_{\rm{min}}$
that is adopted. In this paper, we compute the value of $dE/d\Omega$
from the light-power curves, rather than computing it locally from the
kinetic energy in the flow. As a result, the values of $dE/d\Omega$
that we obtain are smoother 
than those in MLB07 
because they include contributions from
material pointing in different directions with respect to the line of
sight. 
This is similar to the method used in Mizuta \& Aloy (2009) for including contributions from material away from the line of sight.

\begin{figure} 
\plotone{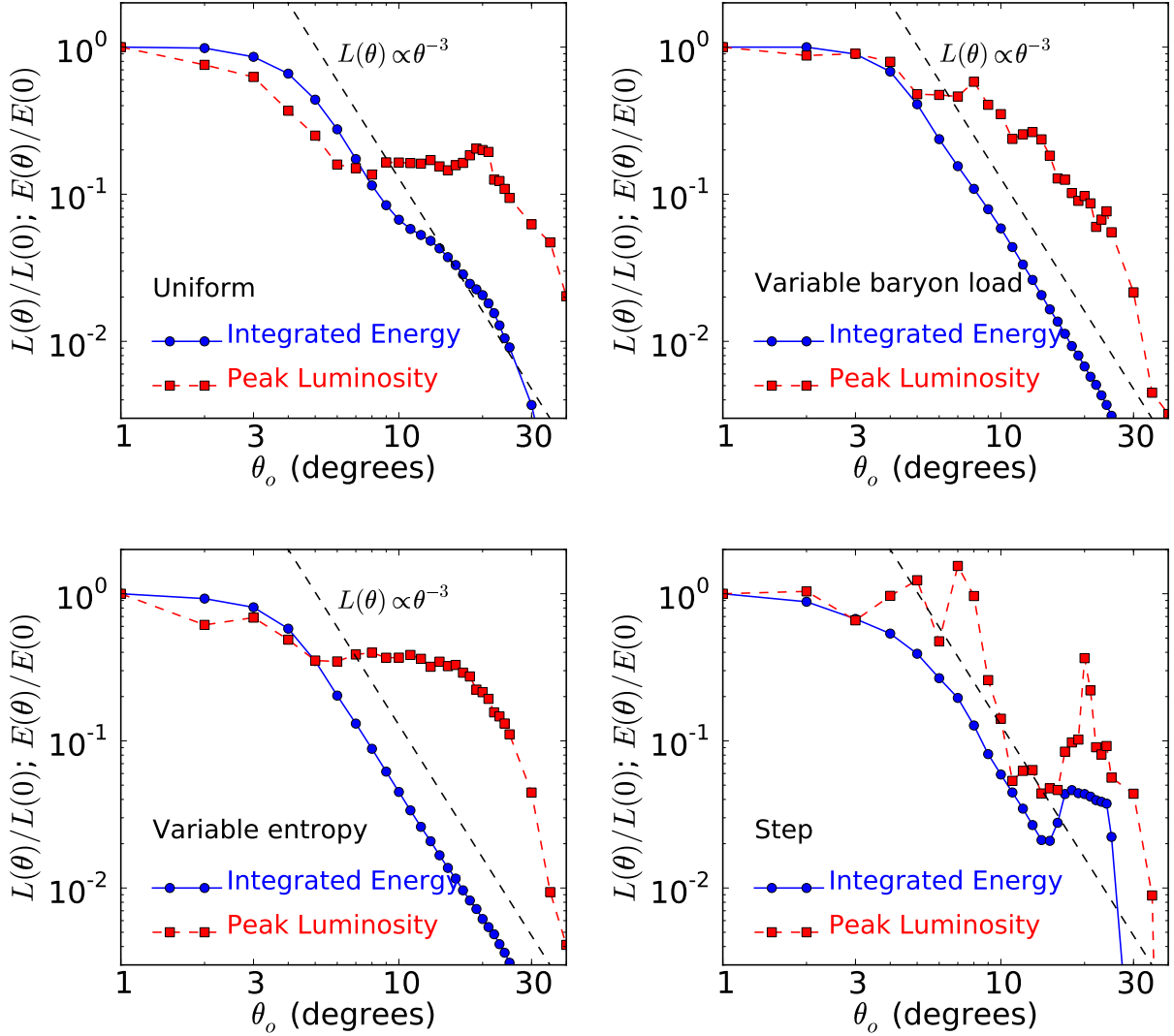} 
\caption{{Angular distribution of the peak luminosity (red dashed
    line) and time integrated energy (blue solid line) measured in the
    light-power curve at $R=2.5\times10^{11}$~cm.}
\label{fig:azi}} 
\end{figure}

Each panel of Figure~\ref{fig:azi} shows the results from one of our
simulations. We find that the isotropic equivalent energy distribution
$dE/d\Omega$ drops with angle smoothly in all cases, with a slope well
represented by a $\theta^{-3}$ power-law, analogously to that found by
Zhang et al. (2004). Only for the {\it step} simulation, a deviation
is observed at $\theta_o\sim20^{\circ}$ off-axis. 
Mizuta \& Aloy (2009) found that the slope ranges from -2.7 for high mass progenitors to -3.7 for low mass progenitors.  However, the progenitor model in their paper most similar to the model we use is model 16OC (from Woosley \& Heger 2006).  For this model, Mizuta \& Aloy (2009) find a 
slope of -2.95, consistent with our results.

Figure~\ref{fig:azi}
also shows the isotropic equivalent peak luminosity distribution
$dL_{\rm{pk}}/d\Omega$, characterized by a much more varied
behavior. At angles less than $\theta_o\sim5^\circ$, the peak
light-power curve tracks the isotropic equivalent energy curve.  At
larger angles, however, it becomes roughly constant, while the
isotropic equivalent energy keeps dropping.  The peak luminosity
remains constant out to an angle that depends on the simulation, but
is roughly a few tens of degrees; it eventually drops for very large
off-axis angles.

\begin{figure} 
\plotone{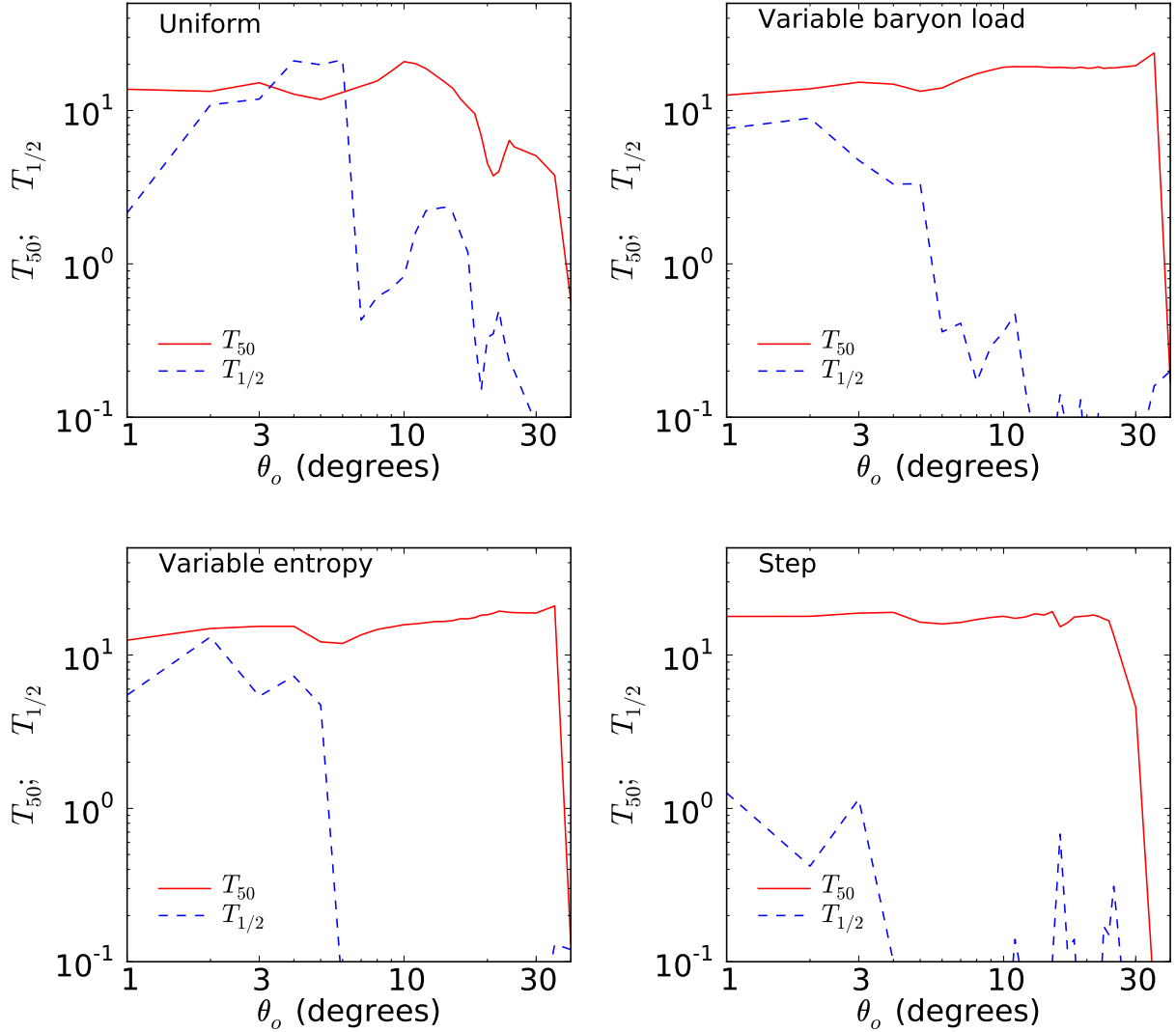} 
\caption{{Duration of the synthetic GRBs as a function of the viewing
    geometry. The solid (dashed) line shows $T_{\rm{50}}$
    ($T_{1/2}$), where $T_{50}$ is the time during which 
    50\% of the total energy is observed, while $T_{1/2}$ is the
    time during which the light-power curve has luminosity greater
    than half the peak luminosity. }
\label{fig:t90}} 
\end{figure}

The angular dependence of the peak luminosity curve is also much more
sensitive than that of the isotropic energy curve to details of the
variability model.  For small off-axis angles ($\theta_o<5^\circ$) ,
increasing the angle decreases the luminosity throughout the duration
of the burst. For larger off-axis angles, however, the loss of total
energy is compensated by the fact that the luminosity is concentrated
into a shorter time (usually the first pulse in the light-power
curve), so that the peak luminosity is kept constant.  To check this,
we computed two measures for the duration of the prompt emission,
$T_{50}$ and $T_{1/2}$. $T_{50}$ is the time during which 50\% of
the total energy is observed, while $T_{1/2}$ time is the time during
which the light curve (in our case the light-power curve) has a
luminosity greater than one half of the peak luminosity. These
quantities are shown in Figure~\ref{fig:t90}. It is interesting to
note that $T_{50}$ and $T_{1/2}$ have very different behaviors,
demonstrating the extent to which the variability properties of the
light-power curves change with observing angle.

\subsection{Power density spectra}

To study the variability of GRB light curves more quantitatively, we
compute the power density spectrum (PDS), i.e., the square modulus of
the Fourier transform of the signal. The PDS of BATSE GRB light curves
is characterized by a power-law slope $PDS(f)\propto f^{-5/3}$ between
cutoffs at $f\sim0.01$~Hz and $f\sim1$~Hz 
(Beloborodov et al. 1998, 2000).

In this section we introduce a new simulation, that we call {\it 
extended uniform}, which is analogous to the {\it uniform}
simulation but has $1/4$th the resolution ($1.25\times10^{8}$~cm) at
the stellar surface and extends to a much larger radius:
$R=2.5\times10^{12}$~cm. This is the simulation used by LMB09 to
derive their on-axis photometric light curve. Here we use the
simulation to compute light-power curves at various off-axis angles at
a radius $R=2.5\times10^{12}$~cm, closer to the radius at which the
radiation is actually released. 
At this radius, the typical spatial resolution is $2\times10^{9}$~cm, although it is allowed to be as high as $5\times10^{8}$~cm.
The temporal resolution of this simulation is $1/15$th of a second, giving a maximum resolved frequency of $7.5$~Hz.
Light-power curves were computed at
off-axis angles $\theta_o=1$, 2, 3, 4, 5, 6, 7, 8, 9, and
$10\degr$. For each of these curves, the power density spectrum was
computed and the 10 spectra were averaged and binned to increase the
signal-to-noise ratio.

The resulting PDS is shown in Figure~\ref{fig:pds1}. The spectrum has
been multiplied by $f^{5/3}$ to emphasize the slope of the power-law
and in an attempt to reproduce, as closely as possible, Fig. 2 of
Beloborodov et al. (1998), whose data are shown as a thin line in the
figure for comparison. The frequencies of Beloborodov et al. (1998)
have been multiplied by a factor 2 to take into account the average
redshift of BATSE GRBs.  The two spectra show a remarkable qualitative
similarity. Both spectra are consistent with a power-law slope
$PDS(f)\propto f^{-5/3}$ between a low- and a high-frequency cutoff.
The locations of the cutoffs are also compatible with observations for
the assumed average redshift of BATSE GRBs.

The rough agreement of the low-frequency cutoff is not surprising
since it is due to the overall duration of the burst, which has been
set in the simulations to be analogous to the average observed GRB
duration. The similarity between the observed and simulated
high-frequency cutoffs is more intriguing, 
although it could be due to either numerical effects or the evolution of the jet.
The time scale associated
with the $\sim3$~Hz cutoff in the synthetic spectra is $\delta
t\sim0.33$~s. The jet at the surface of the star has an opening angle
of $\sim4^\circ$ (see Figure~\ref{fig:oa}) which corresponds to a
transverse size $R_\perp=2.8\times10^9$~cm. For a relativistic sound
speed $c_s=c/\sqrt{3}$, the transverse dimension corresponds to a
crossing time of 0.16~s. 
One possible interpretation is that the cutoff at $\sim3$~Hz can be 
identified as the signature of the jet-crossing time of a disturbance
that propagates from the side to the axis of the jet, such as a 
recollimation shock.
The difference between the $\sim6$~Hz cutoff predicted and the $\sim3$~Hz 
cutoff seen in our simulations could be because either the ``real'' 
opening angle of the jet may be large than the value we use based on 
our definition in Sect. \ref{section3.2}, or the characteristic radius 
at which a shock is crossing the jet may be larger than the original 
stellar radius.
It is also possible that the high-frequency cutoff is a numerical artifact.
The cutoff in our data is at $\sim3$~Hz cutoff, while the maximum 
frequency we can sample in the {\it extended uniform} simulation 
is $7.5$~Hz, close enough that it is possible the cutoff is a result 
of resolution effects in the simulation.
%

The consistency of the observed PDS slope with the synthetic one is,
however, more challenging to explain in terms of the physics of the
jet-star interaction. It is well known that fully developed turbulence
has a 5/3 spectrum (Beloborodov 1998, 2000; Kumar \& Narayan 2009;
Narayan \& Kumar 2009). Turbulent motions are observed in the
simulations, but are concentrated in the cocoon material and not in
the outflow. Even though the collimation shocks are due to the cocoon
pressure on the side of the jet, it is hard to imagine how the
turbulence spectrum can be imprinted directly on the jet light-power
curves.

The qualitative similarity of the observed PDS with the synthetic one is
suggestive that the long time-scale variability in GRBs (down to
tenths of seconds) may be due to the jet interaction with the star and
not to the inner engine variability. However, GRB light curves show
fast spikes with duration of milliseconds or less that cannot be
explained as due to the interaction with the stellar progenitor
(Walker et al. 2000). It seems therefore that there are two sources of
variability in the light curves: the intrinsic variability of the
engine, likely operating on the ms time scale since the engine is a
compact object, and the star-induced variability, with a cutoff at
several tenths of a second, possibly associated with the sound crossing
time of the jet at the star's surface. To study if and how these two
variability sources interfere with each other, we compared the PDS of
the light-power curve from {\it uniform} and {\it variable}
simulations.

Figure~\ref{fig:pds2} shows the comparison of the PDS from the
light-power curves of the various simulations. The power spectrum of
the input variability in the {\it variable} simulations is also
shown. The comparison of the various spectra shows that the two
sources of variability have no significant interaction, with the {\it 
variable} spectra being qualitatively similar to the sum of the {\it 
uniform} spectrum - which contains only the star-induced variability
- and the input spectrum. The two {\it variable} spectra are also
qualitatively similar, indicating that the variability of the jet
depends only on the variability of the total luminosity of the engine
and not on whether the variability is in the entropy or in the baryon
loading of the outflow. 
The PDS of the {\it step} simulation also appears similar to the 
{\it uniform} spectrum, with a series of 
strong resonant peaks at multiples of $5$~Hz, the frequency of the 
on-off cycle, added at higher frequencies.
Again, further simulations will be required to
confirm the generality of the conclusions against different
progenitors and different jet properties.

\section{Discussion and conclusions}

\begin{figure} 
\plotone{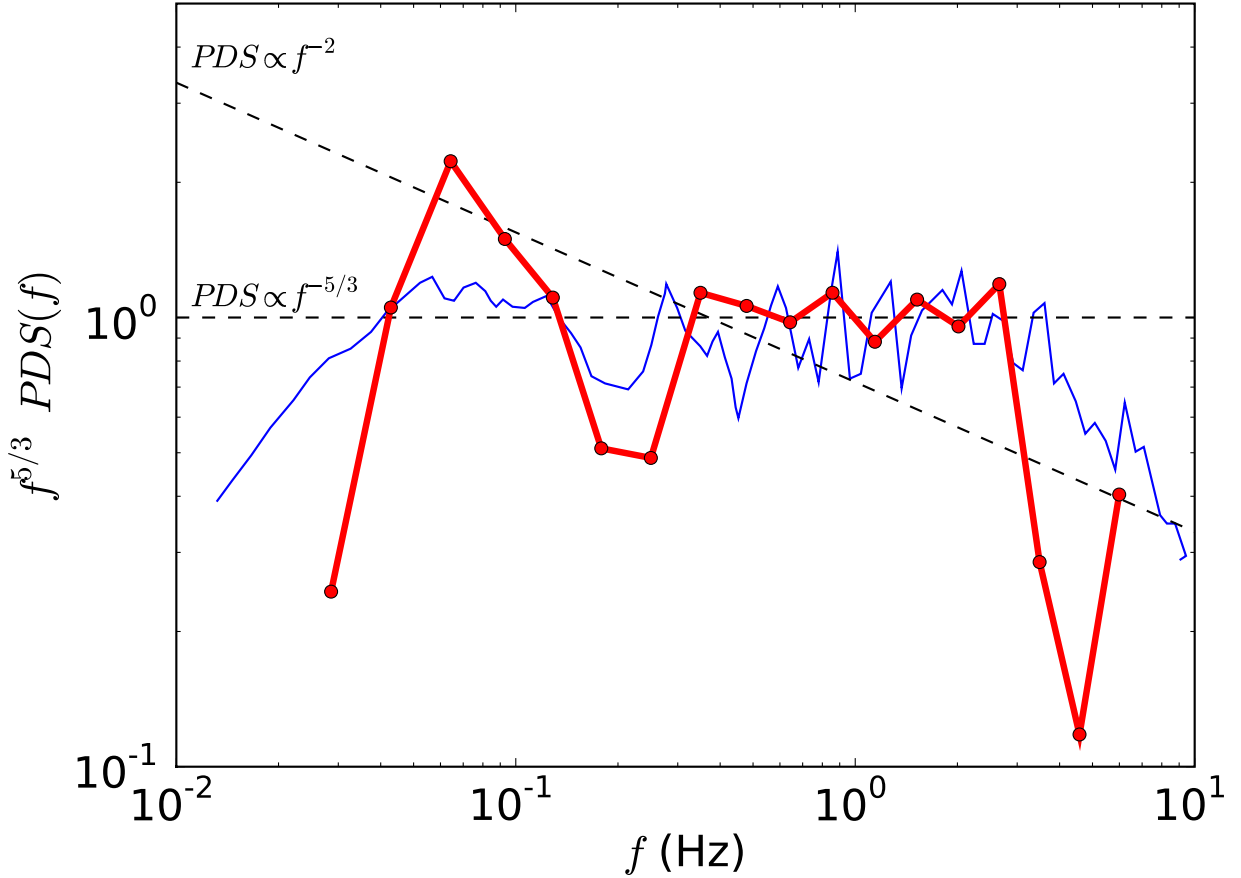} 
\caption{{Average power density spectrum of the light-power curves
    from the {\it extended uniform} simulation computed at a distance
    of $2.5\times10^{12}$~cm from the center of the star. The thick line
    shows the power spectrum obtained by averaging 10 light-power
    curves for viewing angles $\theta_o=1$, 2, 3, 4, 5, 6, 7, 8, 9,
    and $10\degr$ off-axis. The spectrum is multiplied by $f^{5/3}$ to
    reproduce Figure~2 of Beloborodov et al. (1998), whose data are
    shown as a thin line. The dashed lines show
    $PDS(f)\propto{f}^{-5/3}$ and $PDS(f)\propto{f}^{-2}$ spectra for
    comparison.}
\label{fig:pds1}}
\end{figure}

\begin{figure} 
\plotone{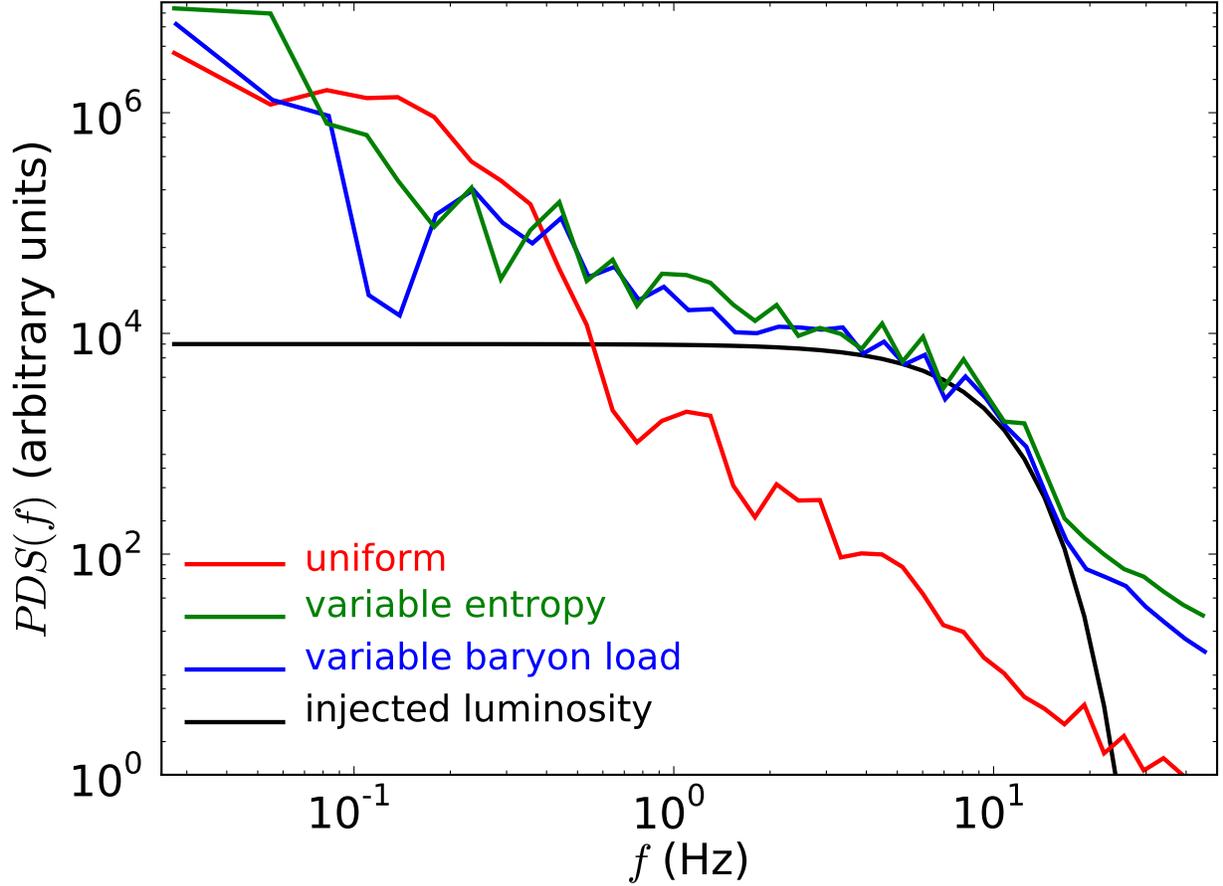} 
\caption{{Comparison of the power spectra of variable simulations with
    the power spectrum from the uniform simulation and the input power
    spectrum at the base of the jet. Even though the spectra are
    plotted in arbitrary units, they have not been normalized and can
    be compared to each other. The power spectra of the {\it variable}
    simulations are similar, showing that the variability properties
    of the light curve do not depend on the way in which the
    variability is injected by the GRB engine. The effect of the
    interaction with the star is clearly present in the PDS curves of
    the {\it variable} simulations as an increase of the power at low
    frequencies, fully compatible with the low-frequency variability
    imprinted on the jet from a uniform engine. The two sources of
    variability add linearly in this case, with the variable
    simulation spectra well described by the sum of the input
    variability spectrum and the uniform jet spectrum.}
\label{fig:pds2}} 
\end{figure}

We have carried out high-resolution hydrodynamic simulations of the
propagation of baryonic GRB jets through their progenitor stars and
beyond. For the first time, we explored the consequences of a variable
injected luminosity with AMR simulations. Our simulations allowed us
to draw a number of conclusions about the role played by the
progenitor star material in shaping the final appearance of the prompt
light curve:

\begin{itemize}

\item The propagation time of the jet in the progenitor star depends
  on the variability properties of the engine in a complex way.  The
  jet propagation inside the star is affected by the detailed
  structure of turbulent eddies inside the cocoon, the development of
  which is highly non-linear and depends on the resolution of the
  simulation. The propagation time is a very important quantity in any
  GRB model since it is the quantity that establishes whether the
  engine is active for a time long enough to power a GRB or not. If
  the engine is active for a time shorter than the propagation time,
  it is more likely that an ``engine powered'' supernova is produced
  (Soderberg et al. 2010) rather than a successful GRB event. The
  propagation time also sets the amount of jet energy that is
  dissipated in the cocoon, available to power a precursor
  (Ramirez-Ruiz et al. 2002; Lazzati \& Begelman 2005) or possibly a
  short duration GRB (Lazzati et al. 2010). High resolution 3D
  simulations are required to nail down this important aspect of the
  jet propagation in the progenitor star.

\item Outside their progenitor stars, jets show a much more
  predictable evolution that does not depend on details.  All jets
  show the three phases of evolution described in MLB07: (i) a
  wide-angle cocoon phase, (ii) a narrow shocked jet phase lasting 10
  to 15 seconds, and (iii) an expanding jet phase. This behavior and
  its timing seem to be fairly independent of the simulation
  resolution as well as the details of the central engine.

\item Contrary to what was previously thought (Zhang et al. 2003;
  MLB07), short time-scale variability injected by the central engine
  is preserved in the jet, even during the shocked phase. The
  variability injected by the central engine is lost only during a
  fast initial phase, lasting a few seconds. On the other hand, the
  interaction of the jet material with the surrounding stellar
  material creates long time-scale variability even for jets injected
  with no variability whatsoever. When a variable jet is injected, we
  find that the progenitor induced variability is not affected, at
  least at the qualitative level. For the range of simulations studied
  here, we find that the properties of the variability at time-scales
  of seconds depend on the progenitor structure and are insensitive to
  the properties of the outflow injected by the central engine. This
  is an important finding because it implies that the fast variability
  in GRB light curves, observable during the bright shocked-jet phase,
  may depend only on the properties of the GRB engine. We can therefore
  study the nature of the GRB engine (still clouded in mystery) by
  studying the properties of the fast variability in GRB light curves.

\item Even though we inject an uniform outflow in the center of the
  star, with no dependence of the jet properties on the polar angle,
  the jet propagating in the circumstellar material is highly
  structured. The total observed energy decreases with the off-axis
  angle as $dE/d\Omega\propto\theta^{-3}$. The peak luminosity, on the
  other hand, is constant between about $5^\circ$ and $20^\circ$. This
  behavior is due to the marked changes in the temporal properties of
  the jet at different off-axis angles. The diversity of light curves
  observed in GRB catalogs can therefore be partly attributed to the
  viewing geometry.

\item We computed the power density spectrum (PDS) of light-power
  curves from an extended simulation (reaching distances from the
  progenitor star ten times larger than our standard simulations). It
  shows two important similarities with the PDS of BATSE light curves
  (Beloborodov et al. 1998; see Fig.~\ref{fig:pds1}): a high frequency
  cutoff at a frequency of few hertz and a power-law
  behavior consistent with a slope $PDS(f)\propto{}f^{-5/3}$ at lower
  frequencies. The high frequency cutoff 
  may tentatively 
  be interpreted as the
  result of the timescale of the propagation of perturbations across
  the jet,
  although the cutoff is near the resolution limit of the simulation 
  and may be a numerical artifact.
  If the presence of the high-frequency cutoff can be confirmed, 
  the coincidence of the simulated cutoff to the
  observed one may be the first direct confirmation that long duration
  GRBs are associated with massive stars at all redshifts and not only
  at $z<1$.  The coincidence between the simulated and observed
  power-law slope suggests instead that the variability that
  characterizes GRB light curves is mostly due to the interaction of
  the jet with the progenitor star and not to the GRB engine. Such a
  result has been obtained under the assumption that the radiation is
  released at a constant radius and at constant efficiency. In
  addition, the PDS is steeper, with a slope $PDS(f)\propto{}f^{-2}$
  if the initial 5 seconds are removed from the light curve. Since the
  structure of the head of the jet is affected by the dimensionality
  of the simulations (cfr. Zhang et al. 2004), we consider this result
  suggestive and not conclusive.

\item A PDS analysis was also performed for the {\it variable}
  simulations (Fig.~\ref{fig:pds2}). The low frequency part of the PDS
  of {\it variable} simulations show a power-law behavior
  qualitatively similar to the one of the {\it uniform}
  simulation. The analogy suggests that, as already discussed in the
  light curve analysis, the long time-scale variability is a result of
  the jet-star interaction that does not depend on the details of the
  engine properties. On the other hand, the high frequency part of the
  {\it variable} PDS has the same qualitative shape of the PDS of the
  luminosity injected by the engine. This suggests that at high
  frequencies the propagation of the jet through the star has no
  impact on the jet structure. The high frequency features are
  transported unmodified by the relativistic outflow. If our results
  are confirmed by simulations with different progenitors and jet
  properties, the analysis of fast variability in GRB light curves
  could give us important clues to the nature of the GRB engine, while
  the analysis of the long time-scale variability can yield important
  clues to the nature of the progenitor star.

\end{itemize}

There are several potential sources the could create variability in the inner $10^{9}$~cm of the star, including instabilities in the propagating jet, instabilities in the region of jet formation, an inherent variability in the energy available from the central engine (i.e. from a variable accretion rate onto a black hole or a varying spin down rate of a proto-magnetar).  Disentangling the sources that give rise to short-timescale variability in observations is likely to be quite challenging.  However, whatever variability does arise in the innermost region of the progenitor is preserved during propagation and remains imprinted on the jet that emerges from the stellar surface.

\begin{figure} 
\plotone{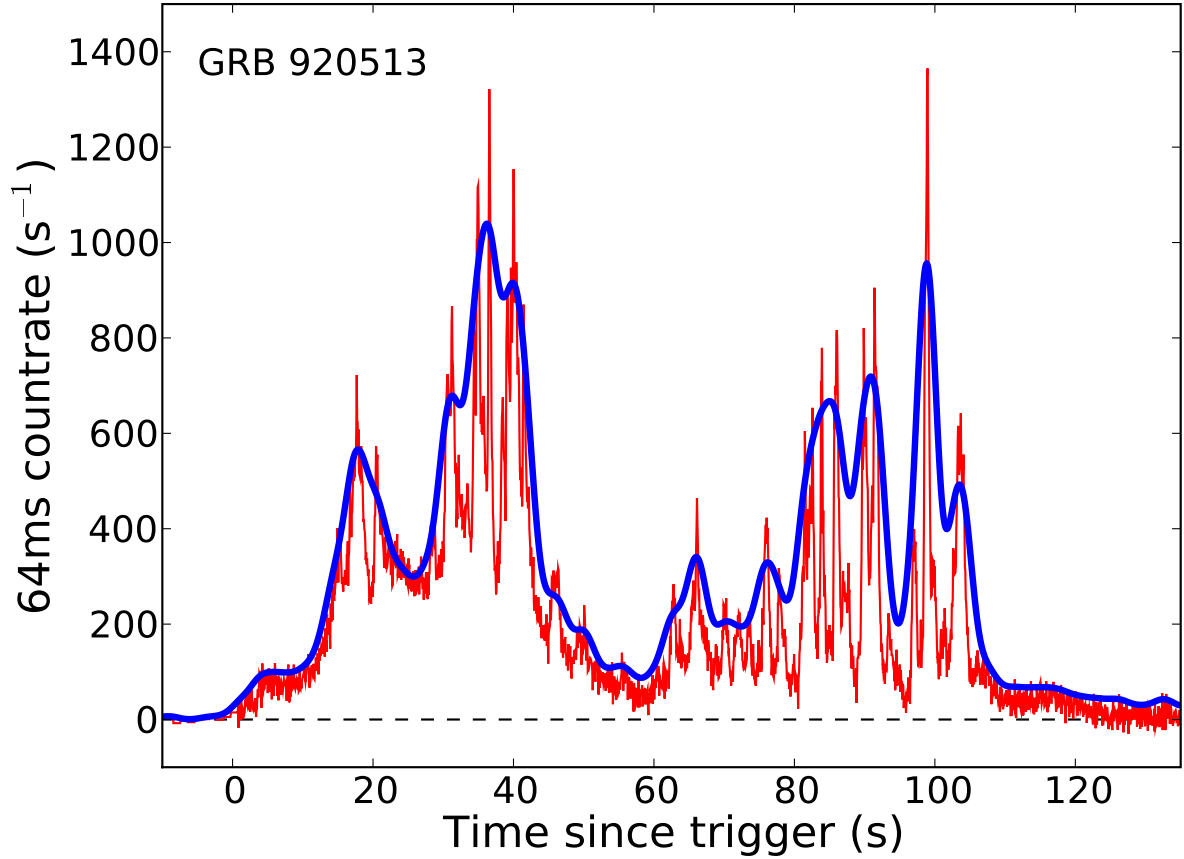} 
\caption{{A cartoon showing the realization of the two sources of
    variability in the light curve of the BATSE GRB~920513. The thin
    line shows the data from BATSE, displaying the short-timescale
    variability due to the activity of the central engine. The thick
    line, on the other hand, emphasizes the long time-scale
    variability envelope imprinted on the jet by the propagation
    through the progenitor star.}
\label{fig:1606}} 
\end{figure}

Based on all the evidence gathered in this project, we advance the
possibility that variability has two origins, with engine and
propagation having comparable roles in shaping the light curve in
different regions of the frequency domain. This possibility is shown
graphically in Figure~\ref{fig:1606} that shows the BATSE light curve
of GRB 920513 with a thick line underlying the envelope of long
time-scale variability due to the jet interaction with the progenitor
star. Faster variability is present in the light curve and that is
interpreted as variability in the luminosity injected by the central
engine at the base of the jet.  Observations of high frequency
variability, therefore, may offer important clues about the nature and
evolution of the central engine.

\acknowledgements The software used in this work was in part developed
by the DoE-supported ASC / Alliance Center for Astrophysical
Thermonuclear Flashes at the University of Chicago. This work was
supported in part by NASA ATP grant NNG06GI06G and Swift GI program
NNX06AB69G and NNX08BA92G. We thank NCSA and the NASA NAS for the
generous allocations of computing time.

%
%
%
%
%
%

\end{document}